# PAMELA: An Open-Source Software Package for Calculating Nonlocal Exact Exchange Effects on Electron Gases in Core-Shell Nanowires

*Andrew W. Long[1] and Bryan M. Wong[2]\**

[1]Department of Materials Science and Engineering, University of Illinois at Urbana Champaign, Urbana, Illinois 61801, USA

[2]Materials Chemistry Department, Sandia National Laboratories, Livermore, California 94551, USA

*Corresponding author. E-mail: usagi@alum.mit.edu

We present a new pseudospectral approach for incorporating many-body, nonlocal exact exchange interactions to understand the formation of electron gases in core-shell nanowires. Our approach is efficiently implemented in the open-source software package PAMELA (Pseudospectral Analysis Method with Exchange & Local Approximations) that can calculate electronic energies, densities, wavefunctions, and band-bending diagrams within a self-consistent Schrödinger-Poisson formalism. The implementation of both local and nonlocal electronic effects using pseudospectral methods is key to PAMELA's efficiency, resulting in significantly reduced computational effort compared to finite-element methods. In contrast to the new nonlocal exchange formalism implemented in this work, we find that the simple, conventional Schrödinger-Poisson approaches commonly used in the literature (1) considerably overestimate the number of occupied electron levels, (2) overdelocalize electrons in nanowires, and (3) significantly underestimate the relative energy separation between electronic subbands. In addition, we perform several calculations in the high-doping regime that show a critical tunneling depth exists in these nanosystems where tunneling from the core-shell interface to the nanowire edge becomes the dominant mechanism of electron gas formation. Finally, in order to present a general-purpose set of tools that both experimentalists and theorists can easily use to predict electron gas formation in core-shell nanowires, we document and provide our efficient and user-friendly PAMELA source code that is freely available at http://alum.mit.edu/www/usagi.

## I. Introduction

The unique electronic properties of semiconductor nanowires continue to be an area of intense interest for diverse applications in optoelectronic, electromechanical, and actuation nanosystems.[1] Because of their small cross sections, quantum confinement effects in nanowires are dominant, and electrons occupy discrete energy levels that are considerably different than the continuum of energy levels found in bulk materials. An area of immense technological interest is the creation of core-shell nanowires (Fig. 1a) where the material composition and/or doping concentration are modulated in the radial direction of the heterostructure nanowire.[2] The core-shell nanowire structure provides a unique advantage compared to homogenous nanowires since it allows confinement of two-dimensional electron gases (2DEGs) at the semiconductor-semiconductor heterojunction interface.[3-9] For the specific case of a GaN/AlGaN core/shell nanowire as shown in Figs. 1a and 1b, carriers are confined within the GaN core by the larger bandgap of the AlGaN shell. In order to optimize performance of these nanowires, it is essential to know which set of material properties (i.e., bandgap alignment, material composition, cross-sectional size, doping density, or many-body electronic effects) favor the spontaneous formation of a mobile electron gas in these nanosystems. Since the experimental parameter space for controlling nanowire properties is immense, theory and predictive modeling play a crucial role in understanding which combination of material parameters can be used to optimize performance in these nanoscale systems.

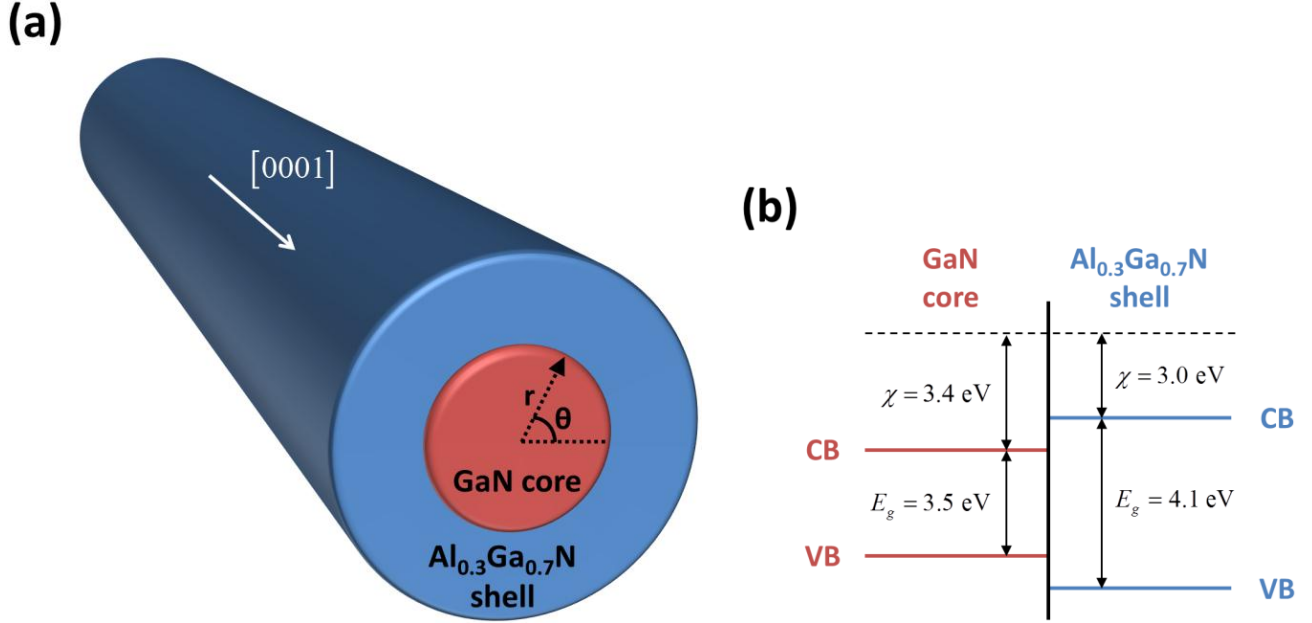

**Figure 1.** (a) Schematic of the cylindrical GaN/AlGaN core/shell nanowires considered in this work. The nanowire has its axis in the [0001] direction. (b) Valence band (VB) and conduction band (CB) alignment at the core/shell interface. For the particular GaN/Al$_{0.3}$Ga$_{0.7}$N composition, the band alignment leads to a Type I (straddling) heterojunction with a $\Delta E_c$ = 0.4 eV conduction band discontinuity.

The most commonly used approach for predicting electronic properties in these nanowires is a simplistic Schrödinger-Poisson formalism[10-12] where a coupled system of Schrödinger and Poisson equations are solved self-consistently to obtain band-bending diagrams, wavefunctions, and electron densities. In this simple approach, an effective-mass Schrödinger equation of the form

$$\left[-\frac{\hbar^2}{2}\nabla_i \cdot \frac{1}{m^*(r_i)}\nabla_i + V_n(\vec{r}_i) + V_{\text{CB}}(\vec{r}_i) + V_{\text{e-e}}(\vec{r}_i)\right]\psi_i(\vec{r}_i) = E_i\psi_i(\vec{r}_i), \quad (1)$$

is solved, where $m^*$ is the effective mass, $\psi_i$ is the electron wavefunction for state $i$, $E_i$ is its energy, $V_n$ is the spatially dependent free carrier concentration due to the *n*-type doping density, $V_{CB}$ is the conduction band edge, and $V_{e-e}$ is the (local) Poisson potential due to electron-electron interactions. Within the conventional Schrödinger-Poisson formalism, the Poisson potential, $V_{e-e}$, is given by

$$V_{e-e}(\vec{r}_i) = \sum_j \int d\vec{r}_j \, \frac{2e^2 |\psi_j(\vec{r}_j)|^2}{4\pi\varepsilon\varepsilon_0 |\vec{r}_i - \vec{r}_j|} N(E_F, E_j, T), \quad (2)$$

where $N(E_F, E_j, T)$ is the integrated one-dimensional density of states along the nanowire, and the summation in Eq. (2) is over *all occupied orbitals*, $\psi_j$. However, since the summation is over all $j$, an electron feels the electrostatic potential from the other electrons *as well as from itself*. This unphysical *self-interaction error* is well-known from electronic structure theory and catastrophically leads to overdelocalized electrons, underestimated band gaps, and unstable charge states.[13-17] Instead, a more accurate (but still not quite correct) expression for $V_{e-e}$ is

$$V_{e-e}(\vec{r}_i) = \sum_{j \neq i} \int d\vec{r}_j \, \frac{2e^2 |\psi_j(\vec{r}_j)|^2}{4\pi\varepsilon\varepsilon_0 |\vec{r}_i - \vec{r}_j|} N(E_F, E_j, T), \quad (3)$$

where $V_{e-e}$ is now an *orbital-dependent* potential. In other words, the difference between Eq. (2) and (3) is the exclusion of the "diagonal term," $j = i$, which can be interpreted as the interaction of the electron with itself. However, the use of Eq. (3) (which, again, is still not quite correct as it ignores antisymmetry of the electronic wavefunction) poses severe computational difficulties in the conventional Schrödinger-Poisson approach since this orbital-dependent operator is *different* for every orbital (because of the restricted summation over $j \neq i$), and one would have to solve a different effective Schrödinger equation *for each individual orbital* in the self-consistent approach.

Herein, we provide a simple and efficient approach for rigorously eliminating self-interaction errors and assessing the effects of many-body, electronic interactions on the formation of electron gases in core-shell nanowires. Our approach utilizes a pseudospectral numerical method[18,19] for calculating many-body, nonlocal exact exchange interactions within a nanowire and is incorporated in the open-source software package PAMELA (Pseudospectral Analysis Method with Exchange & Local Approximations) available online.[20] We specifically focus on the effects of nonlocal exact exchange in nanowire systems for three important reasons: (1) Very recent work by us and others have shown that electron gas formation is insensitive to *local* correlation effects.[11,12] While results obtained with a simplistic local density approximation (LDA) were nearly identical to uncorrelated calculations, the effects of nonlocal exact exchange on large core-shell nanowires are not known; (2) We and other researchers have also previously shown that exchange-correlation effects are inherently nonlocal, and electronic properties in molecules and bulk systems can be significantly improved with nonlocal exchange,[21-28] and (3) Electronic wavefunctions with exact exchange are usually the initial reference state for including higher-level treatments of electron correlation,[29-32] so the formalism used in this work serves as a starting point for even more extensive treatments of strong correlation effects in nanowires. While the incorporation of nonlocal exchange in atomistic calculations of molecules and materials is well-known, we are unaware of any previous work assessing the effect of exact exchange on large core-shell nanowires.

The work presented in this paper is divided into two separate parts: In the first part (Sections I-IV), we give a brief derivation of the exact-exchange equations for electrons in a core-shell nanowire, and we demonstrate how to reduce the integro-differential exact-exchange equations to simultaneous (coupled) partial differential equations that can be more easily solved numerically. Using this approach, we present specific situations where exact exchange effects become dominant and give several examples showing electron-occupancy levels, band-bending diagrams, electron density delocalization patterns, and electron subband energies. In the second part of the paper (Appendix Sections 1-9), we give a more detailed description of the various approaches and numerical routines which comprise the PAMELA software package that is available online.[20] The availability of the efficient and user-friendly PAMELA source code (< 400 lines, including several lines of comments) is intended to reduce the technical burden of writing computer code from scratch and to present a general-purpose set of tools that both experimentalists and theorists can easily use to predict electron gas formation in core-shell nanowires. With these new approaches and predictive methods, we highlight several areas where many-body quantum effects play a significant role in both the formation and distribution of electrons in these low-dimensional nanostructures.

**II. Theory and Methodology**

The Hamiltonian for *N* electrons in a nanowire is given by

$$\hat{H} = \sum_{i=1}^{N} \hat{h}_i + \hat{V}_n(\vec{r}_i) + \hat{V}_{CB}(\vec{r}_i) + \sum_{j \neq i}^{N} \frac{e^2}{4\pi\varepsilon\varepsilon_0} \frac{1}{|\vec{r}_i - \vec{r}_j|}, \quad (4)$$

where

$$\hat{h}_i = -\frac{\hbar^2}{2} \nabla_i \cdot \frac{1}{m^*(r_i)} \nabla_i, \quad (5)$$

$\hat{V}_n$ is the spatially dependent free carrier concentration due to the *n*-type doping density, and $\hat{V}_{CB}$ is the conduction band edge. To satisfy the Pauli exclusion principle, we use a fully antisymmetric wavefunction given by

$$\begin{aligned}\Psi &= A_N\left(\psi_1, \bar{\psi}_1, \psi_2, \bar{\psi}_2, \ldots, \psi_{N/2}, \bar{\psi}_{N/2}\right) \\ &= \left|\psi_1, \bar{\psi}_1, \psi_2, \bar{\psi}_2, \ldots, \psi_{N/2}, \bar{\psi}_{N/2}\right|,\end{aligned} \quad (6)$$

where an overbar denotes that a spatial orbital, $\bar{\psi}_i$, has a spin-down electron, and the lack of a bar, $\psi_i$, denotes that a spatial orbital has a spin-up electron. $A_N$ is the antisymmetrization operator and can be represented as a single Slater determinant. Using this antisymmetric wavefunction, the generalized Hartree-Fock equations[33] for a nanowire reduce to

$$\left[\hat{h}_i + \hat{V}_n(\vec{r}_i) + \hat{V}_{CB}(\vec{r}_i) + \sum_{j=1}^{N} \hat{V}_{D,j}(\vec{r}_i, \vec{r}_j) - \sum_{j=1}^{N} \hat{V}_{EXX,j}(\vec{r}_i, \vec{r}_j)\right] \psi_i(\vec{r}_i) = E_i \psi_i(\vec{r}_i), \quad (7)$$

where $\hat{V}_D$ is the direct interaction:

$$\hat{V}_{D,j}(\vec{r}_i,\vec{r}_j)\psi_i(\vec{r}_i) = \left[\int d\vec{r}_j\ \psi_j^*(\vec{r}_j)\left(\frac{e^2}{4\pi\varepsilon\varepsilon_0}\frac{1}{|\vec{r}_i-\vec{r}_j|}\right)\psi_j(\vec{r}_j)N(E_F,E_j,T)\right]\psi_i(\vec{r}_i), \quad (8)$$

and $\hat{V}_{EXX}$ is the (nonlocal) exchange interaction:

$$\hat{V}_{EXX,j}(\vec{r}_i,\vec{r}_j)\psi_i(\vec{r}_i) = \left[\int d\vec{r}_j\ \psi_j^*(\vec{r}_j)\left(\frac{e^2}{4\pi\varepsilon\varepsilon_0}\frac{1}{|\vec{r}_i-\vec{r}_j|}\right)\psi_i(\vec{r}_j)N(E_F,E_j,T)\right]\psi_j(\vec{r}_i). \quad (9)$$

In Eqs. (8) and (9), $N(E_F,E_j,T)$ is the integrated electron concentration given by

$$N(E_F,E_j,T) = \int dE\ g(E,E_j)f(E,E_F,T), \quad (10)$$

where $g(E,E_j)$ is the one-dimensional density of states, $E_j$ is the energy of the $j$th wavefunction, $E_F$ is the Fermi energy, $T$ is the temperature, and $f(E,E_F,T)$ is the Fermi-Dirac distribution (all described further in Section 4 of the Appendix). Note that the exchange interaction is a nonlocal operator (where $\psi_i$ is inside the integral) and involves an "exchange" of electron $i$ and $j$ to the right of $1/|\vec{r}_i-\vec{r}_j|$ in Eq. (9) relative to Eq. (8). Also note that the summation in Eq. (7) is unrestricted and includes terms where $i=j$. The unrestricted summation is a unique property of using an antisymmetric wavefunction since the direct and exchange interactions for a single electron are equal but opposite in sign, and the self-interaction cancels out exactly[33] (this is easily seen by inspecting Eqs. (8) and (9), and showing that $\left[\hat{V}_{D,i}(\vec{r}_i,\vec{r}_i) - \hat{V}_{EXX,i}(\vec{r}_i,\vec{r}_i)\right]\psi_i(\vec{r}_i) = 0$). If we use an antisymmetric wavefunction where each of the occupied spatial orbitals is doubly occupied as implied in Eq. (6), we can further simplify Eq. (7) to a sum over the number of doubly-occupied spatial orbitals[33]:

$$\left[\hat{h}_i + \hat{V}_n(\vec{r}_i) + \hat{V}_{CB}(\vec{r}_i) + \sum_{j=1}^{N/2} 2\hat{V}_{D,j}(\vec{r}_i,\vec{r}_j) - \sum_{j=1}^{N/2}\hat{V}_{EXX,j}(\vec{r}_i,\vec{r}_j)\right]\psi_i(\vec{r}_i) = E_i\psi_i(\vec{r}_i), \quad (11)$$

Eq. (8) is a nonlinear integro-differential equation that involves both a local $\hat{V}_{D,j}$ operator and a nonlocal integral operator due to $\hat{V}_{EXX}$. Examining first the integral representing the direct interaction between electrons:

$$2V_{D,j} = 2\int d\vec{r}_j \ \psi_j^*(\vec{r}_j)\left(\frac{e^2}{4\pi\varepsilon\varepsilon_0}\frac{1}{|\vec{r}_i-\vec{r}_j|}\right)\psi_j(\vec{r}_j)N(E_F,E_j,T), \tag{12}$$

and acting on $V_{D,j}$ with the operator $\nabla_i^2$ (which depends only on $\vec{r}_i$) gives

$$\begin{aligned}\nabla_i^2 2V_{D,j} &= 2\int d\vec{r}_j \ \psi_j^*(\vec{r}_j)\nabla_i^2\left(\frac{e^2}{4\pi\varepsilon\varepsilon_0}\frac{1}{|\vec{r}_i-\vec{r}_j|}\right)\psi_j(\vec{r}_j)N(E_F,E_j,T) \\ &= 2\int d\vec{r}_j \ \psi_j^*(\vec{r}_j)\left(\frac{-e^2\delta^3(\vec{r}_i-\vec{r}_j)}{\varepsilon\varepsilon_0}\right)\psi_j(\vec{r}_j)N(E_F,E_j,T) \\ &= -\frac{2e^2}{\varepsilon\varepsilon_0}|\psi_j(\vec{r}_j)|^2 N(E_F,E_j,T).\end{aligned} \tag{13}$$

where we have made use of the identity $\nabla_i^2 \frac{1}{|\vec{r}_i-\vec{r}_j|} = -4\pi\delta^3(\vec{r}_i-\vec{r}_j)$; in other words, the Green's function for Laplace's equation is $-1/(4\pi|\vec{r}_i-\vec{r}_j|)$. One can immediately recognize that Eq. (13) is simply Poisson's equation for electron charge densities, which is the expression used in conventional Schrödinger-Poisson approaches.[10-12]

Using the same techniques as in the treatment of the direct interaction term,[19,34,35] the exchange interaction between electrons is

$$V_{EXX,j} = \int d\vec{r}_j \ \psi_j^*(\vec{r}_j)\left(\frac{e^2}{4\pi\varepsilon\varepsilon_0}\frac{1}{|\vec{r}_i-\vec{r}_j|}\right)\psi_i(\vec{r}_j)N(E_F,E_j,T), \tag{14}$$

and acting on $V_{EXX,j}$ with the operator $\nabla_i^2$ (which depends only on $\vec{r}_i$) gives

$$\begin{aligned}\nabla_i^2 V_{EXX,j} &= \int d\vec{r}_j \ \psi_j^*(\vec{r}_j)\nabla_i^2\left(\frac{e^2}{4\pi\varepsilon\varepsilon_0}\frac{1}{|\vec{r}_i-\vec{r}_j|}\right)\psi_i(\vec{r}_j)N(E_F,E_j,T) \\ &= -\frac{e^2}{\varepsilon\varepsilon_0}\psi_j^*(\vec{r}_i)\psi_i(\vec{r}_i)N(E_F,E_j,T)\end{aligned} \tag{15}$$

Finally, the potential energy due to the *n*-type doping density is given by solving

$$\nabla\cdot\varepsilon^*\varepsilon_0\nabla V_n(\vec{r}_i) = |e|\rho_D(\vec{r}_i), \tag{16}$$

where $\rho_D$ is the charge density arising from the ionized dopants. It is important to note that we have derived an approach where the integro-differential exact-exchange equations have been reduced to simultaneous (coupled) partial differential equations *without any nonlocal integral terms*.

Eqs. (11), (13), (15), and (16) comprise a set of Schrödinger, direct, exchange, and Poisson equations that are non-linearly coupled with each other and must be solved iteratively to self-consistency. We obtain solutions and electronic properties (energies, densities, wavefunctions, and band-bending diagrams) from these coupled equations by using a pseudospectral approach implemented in our PAMELA open-source code,[20] described in further detail in the Appendix. Eqs. (11), (13), (15), and (16) are defined by appropriate boundary conditions and constraints. At the surface of the cylindrical nanowire, we set $\psi_i(\vec{r} = r_{shell}) = 0$ (there is no "leakage" of electrons outside the nanowire edge, $r_{shell}$), and we set $V_D(\vec{r} = r_{shell}) = V_{EXX}(\vec{r} = r_{shell}) = V_n(\vec{r} = r_{shell}) = V_{CB}(\vec{r} = r_{shell}) = 0$ (these conditions define our zero reference of energy to be at the $\vec{r} = r_{shell}$ conduction band edge). In addition, we determine the position of the Fermi level via a charge-neutrality condition (described in Section 4 of the Appendix) that adjusts the Fermi level at each iteration step.

The specific material system we consider in this study is a cylindrical nanowire with a GaN core and an $Al_{0.3}Ga_{0.7}N$ shell. We assume the GaN/$Al_{0.3}Ga_{0.7}N$ layers are in the Wurtzite structure oriented with the [0001] direction along the axis of the nanowire (see Fig. 1a), which we also assume to be defect-free.[5] While we focus on this particular material system, we expect our findings to hold for other similar core-shell nanowires with Type I (straddling) heterojunctions. The bulk electronic properties and material parameters used in our self-consistent calculations are given in Table 1.

|  | GaN | Al$_{0.3}$Ga$_{0.7}$N |
|---|---|---|
| $E_g$ (eV) | 3.475 | 4.092 |
| $\chi$ | 3.448 | 3.016 |
| $\Delta E_c$ (eV) | 0.432 | |
| $m^*$ | 0.40 | 0.236 |
| $\varepsilon_r$ | 9.28 | 9.097 |

**Table 1.** Material parameters used in our self-consistent calculations. $E_g$ is the band gap, $\chi$ is the affinity, $\Delta E_c$ is the conduction band offset, $m^*$ is the relative electron effective mass, and $\varepsilon_r$ is the relative static dielectric constant ($\chi$ is taken from Ref. 11, and other parameters are from Ref. 36).

**III. Results and Discussion**

Numerous simulations with the PAMELA program[20] were carried out on cylindrical core-shell nanowires with GaN core radii ranging from 10 nm to 50 nm with a fixed 20-nm-thick shell layer of Al$_{0.3}$Ga$_{0.7}$N. We investigated $N$-type doping densities ranging from $10^{16}$ to $1.2\times10^{18}$ cm$^{-3}$, assuming a uniform doping density throughout the nanowire. In order to numerically solve the coupled pseudospectral differential equations, we chose 79 Chebyshev points to discretize the radial direction, and 24 equispaced points were chosen for the $\theta$ discretization (see Sections 6-9 of the Appendix), resulting in matrices of rank 936 used throughout our calculations.

We first discuss the effects of nonlocal exact exchange on the number of occupied wavefunctions in our nanowire, as demonstrated in the contour plots in Fig. 2. Calculations without exact exchange consistently overestimate the number of occupied levels, which become even more noticeable at higher doping densities and at higher core radii. At smaller doping densities and small core-radii, exact exchange has a noticeably larger effect on the electron density profile within the cross section of the nanowire, as shown in Figs. 3a-d. For a small nanowire with a 20-nm-wide core, a 20-nm-thick shell, and a doping density of $n_D = 0.5 \times 10^{17}$ cm$^{-3}$, we find that the electrons stay localized at the center of the core when nonlocal exact exchange effects are incorporated. In contrast, we find that calculations without exchange (using the same doping density as before) predict a charge density that is more delocalized, showing a pronounced annular distribution. In particular, we obtain an electron density with a peak height that is reduced by almost half when exchange interactions are not incorporated.

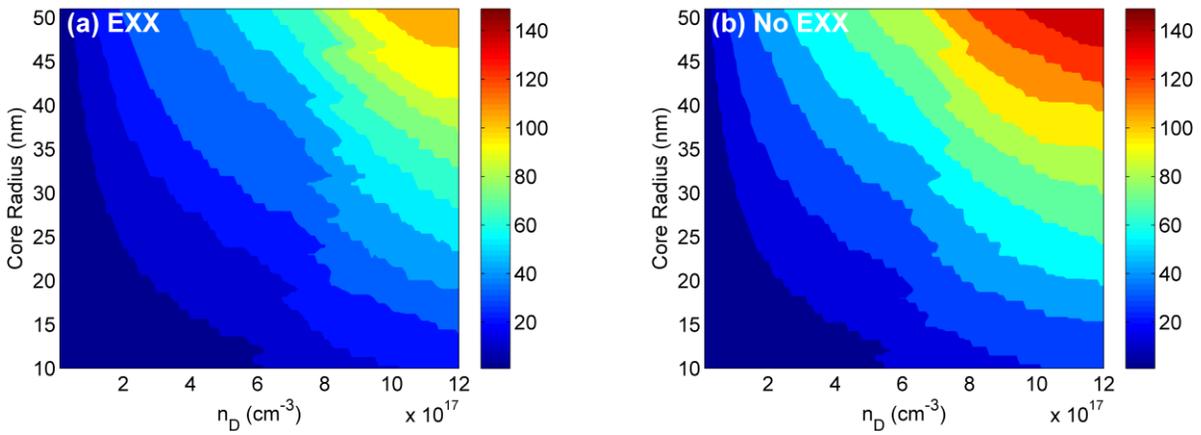

**Figure 2.** Contour plots showing the electron occupancy number in a nanowire as a function of core radius and doping density, $n_D$, (a) with exact exchange and (b) without exact exchange.

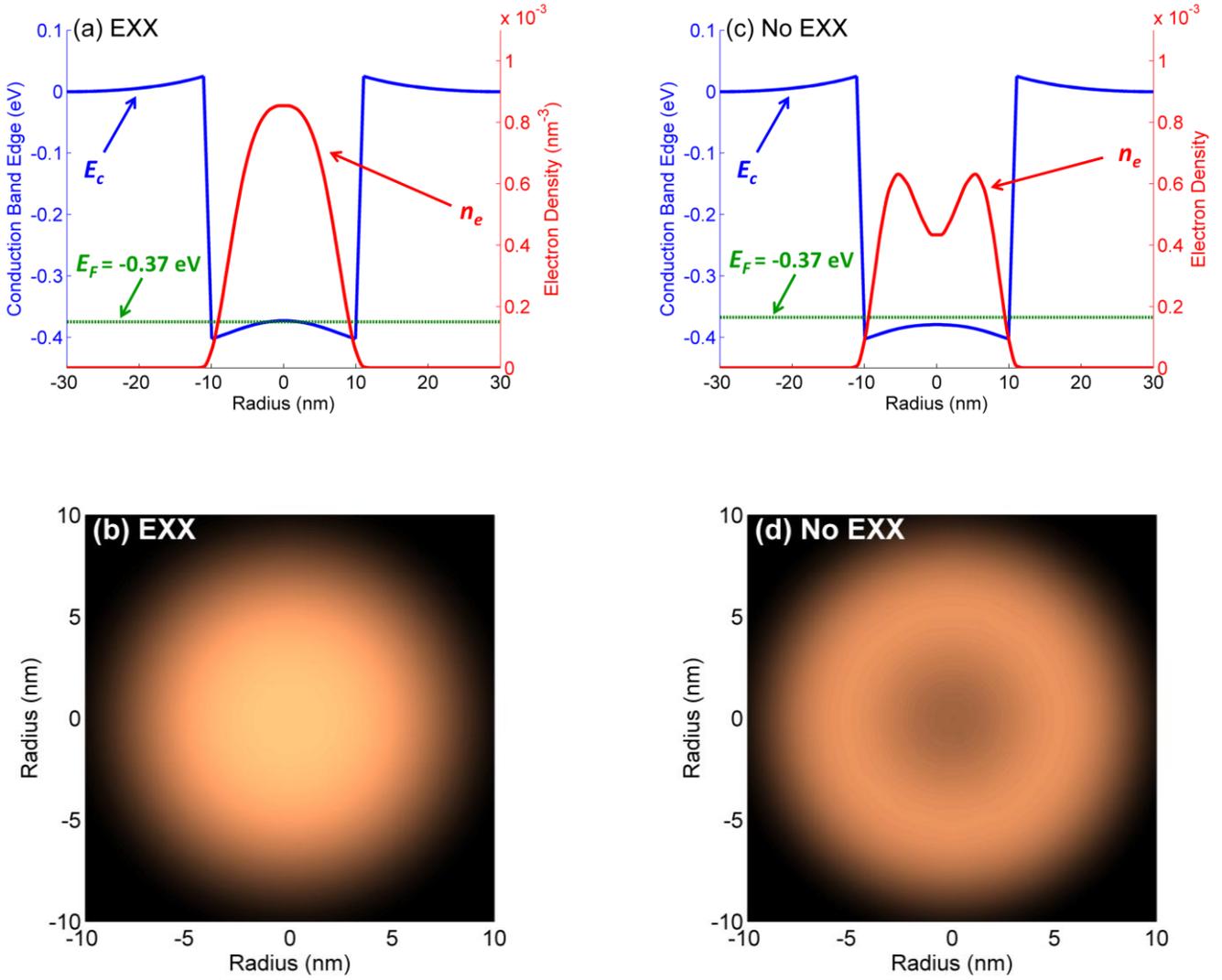

**Figure 3.** (a) Band-bending (blue, left axis), Fermi energy (green), and electron density (red, right axis) for a nanowire with a 20-nm-wide core and a 20-nm-thick shell, and (b) cross-sectional distribution of the free electron gas for calculations incorporating nonlocal exact exchange. Panels (c) and (d) show the corresponding results for simulations without exact exchange. For both the exact exchange and no exchange cases, we used the same doping density of $n_D = 0.5 \times 10^{17} cm^{-3}$.

We see similar localization effects in nanowires with larger core sizes and higher doping densities. In Figs. 4a and b, we use the PAMELA program to plot the free electron gas distribution for nanowires with an 80-nm-wide core, a 20-nm-thick shell, and a doping density of $n_D = 5.5 \times 10^{17}$ cm$^{-3}$. In both cases we see small variations in the electron density in the middle of the core region; however, the localization of the electron density at the core-shell interface is sharper with exact exchange, whereas the electron gas without exact exchange decays slowly towards the nanowire center.

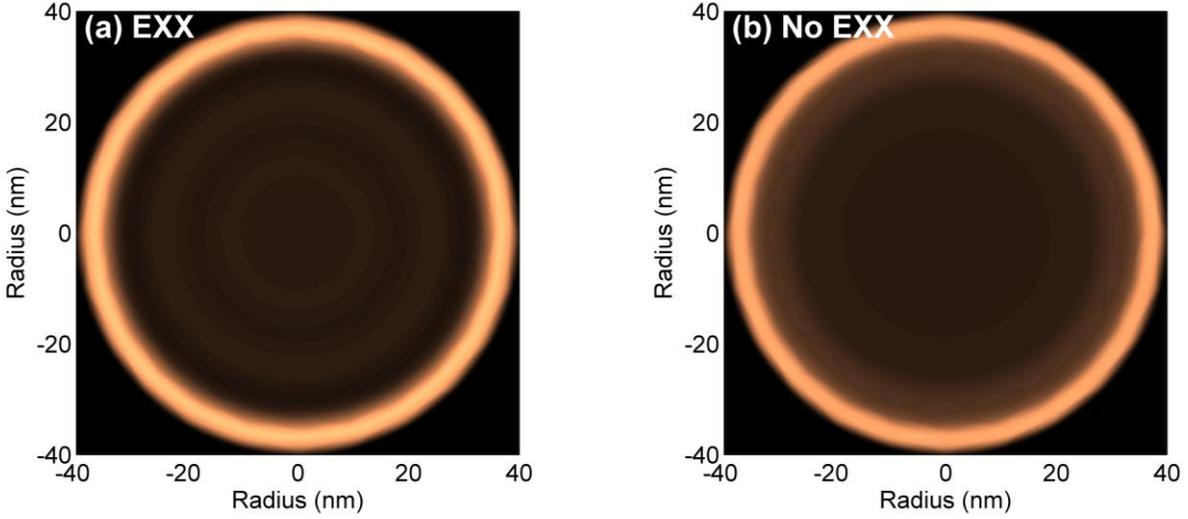

**Figure 4.** Cross-sectional distribution of the free electron gas for a nanowire with an 80-nm-wide core and a 20-nm-thick shell for self-consistent calculations (a) with exact exchange and (b) without exact exchange. The doping density used for both cases is $n_D = 5.5 \times 10^{17} \text{cm}^{-3}$. With exact exchange, the free electron gas is more localized at the heterojunction interface, whereas the electron gas without exact exchange decays slowly towards the nanowire center.

In order to explain the electron delocalization patterns in Figs. 3-4, we must take a closer look at the effective electron-electron potential energy in a nanowire. As derived in Eq. (11), the effective potential energy due to electron-electron repulsion in a nanowire is given by a difference of two terms: $\sum_{j=1}^{N/2} \left( 2\hat{V}_{D,j} - \hat{V}_{EXX,j} \right)$, where $\hat{V}_{D,j}$ represents the direct Coulombic repulsion between electrons, and $\hat{V}_{EXX,j}$ is the nonlocal quantum-mechanical exchange potential. In particular, one can view the nonlocal exchange potential as a "correction" to the local Coulomb potential to take into account the antisymmetry of the electronic wavefunction. In conventional Schrödinger-Poisson approaches that *do not* incorporate exchange,[10] the bare $\hat{V}_{D,j}$ term *exaggerates* the effect of Coulombic repulsion between electrons, leading to electronic densities that are more spread out, or delocalized (i.e., negatively-charged electrons tend to avoid each other). In contrast, the exchange potential, $\hat{V}_{EXX,j}$, is a negative contribution to the effective potential energy (see Eq. (11)) and compensates for the exaggeration due to the $\hat{V}_{D,j}$ term, leading to a more localized electronic density. As a result, the inclusion of quantum-mechanical exchange plays a crucial role in correcting the over-delocalization of electrons in nanowire systems that arises from including only the bare Coulombic interaction.

We can further analyze the effects of exact exchange on the energy eigenvalues (or quasiparticle energies) of the electronic subbands outputted by the PAMELA program. As shown in Fig. 5, we plot the lowest 20 electron subband eigenvalues as a function of doping density for calculations with and without nonlocal exact exchange. Due to the cylindrical symmetry of the nanowire, some eigenvalues are either singly or doubly degenerate and are shown as blue or red lines, respectively. In general we see that the energy spacings between subbands are smaller in the absence of exchange, while exchange effects dramatically increase these energy separations (i.e., the subbands are more spread out in Fig. 5a). This widening of energy spacings is consistent with many-body electronic effects where it is well-known that nonlocal exact exchange widens quasiparticle gaps in both molecules and bulk materials.[23,24,26] In contrast, this nonlocal many-body electronic effect is completely absent when solving the conventional Schrödinger-Poisson equations without exchange, leading to energy differences between subbands that are considerably underestimated. It is also important to note that the relative energy spacings between subbands further explain the overestimation in the number of occupied levels shown in Fig. 2b. Specifically, the total electron density, $n_e$, is proportional to the difference between the Fermi energy and the occupied electronic subbands: $n_e \propto \sum_{j=1}^{N/2} \sqrt{E_F - E_j}$ (see Eq. A24). However, in order to satisfy the charge neutrality constraint, we require the total number of positive and negative charges over the entire nanowire to be equal: $\int dA\, n_D = \int dA\, n_e$ (where $A$ is the cross-sectional area; see Section 4 of the Appendix). If the energy separation between subbands is small (as it is for the case without exchange), then the various $E_F - E_j$ terms will also be small, and we will require considerably more occupied levels to satisfy the charge neutrality constraint. For the same reason, since the energy spacings between subbands are considerably larger when exact exchange is incorporated, the $E_F - E_j$ terms are also larger, and we do not need as many occupied levels to satisfy charge neutrality (cf. Fig. 2a).

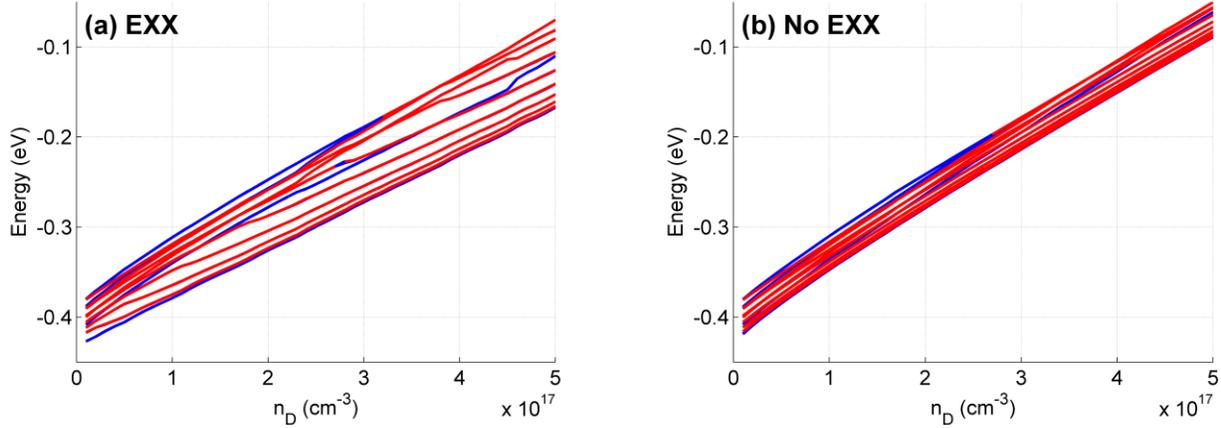

**Figure 5.** Lowest 20 electron subband eigenvalues as a function of doping density for calculations (a) with nonlocal exact exchange and (b) without exact exchange. The simulations in both (a) and (b) are for a nanowire with a 20-nm-wide core and a 20-nm-thick shell. The red lines indicate eigenvalues that are doubly degenerate.

Finally, in order to quantify the effects of nonlocal exact exchange on the spatial localization of electrons in our nanowires, we introduce a dimensionless core-normalized variance of the electron density, defined as

$$\langle r^2 \rangle = \frac{\int dA \, r^2 n_e(r,\theta)}{r_{core}^2 \int dA \, n_e(r,\theta)}, \qquad (17)$$

where $A$ is the cross-sectional area of the nanowire. By construction, $\langle r^2 \rangle$ measures the relative spread of the electron density and is bounded between 0 and $r_{shell}^2 / r_{core}^2$. From its definition, values of $\langle r^2 \rangle$ less than 1 indicate localization within the core, and values greater than 1 represent confinement within the shell layer. In Fig. 6a, we plot the core-normalized variance as a function of doping density for a set of representative core radii. At low doping density ($< 2\times10^{17}$ cm$^{-3}$), we find that $\langle r^2 \rangle$ rises rapidly and eventually saturates to ~0.7 at moderate doping. This variation in $\langle r^2 \rangle$ corresponds to a shift in electron localization from the center of the nanowire to a spreading of electron density within the core as the doping density is increased. At a critical doping density (~ $7\times10^{17}$ cm$^{-3}$ for these particular core radii), Fig. 6a shows a dramatic, abrupt change as $\langle r^2 \rangle$ rapidly increases towards 1. This sudden change in behavior corresponds to a transition where the electron gas in the middle of the core becomes strongly localized at (or even beyond) the heterojunction core-shell interface. It is important to note that these trends for $\langle r^2 \rangle$ still persist whether exact exchange is incorporated or not; however, the transition to an interface-localized state occurs at higher doping densities when exact exchange is included. This interesting behavior in $\langle r^2 \rangle$ has a direct correspondence with the position of the Fermi level in the core-shell nanowire. Fig. 6b shows the Fermi energy as a function of doping density for the same set of core radii considered previously. As the doping density is increased, the Fermi level rises linearly as we require a larger number of occupied electrons to satisfy charge neutrality. However, at exactly the same critical doping densities shown in Fig. 6a, we find that the Fermi energy in Fig. 6b has reached a saturation point such that further increases in doping density do not dramatically affect the position of the Fermi level.

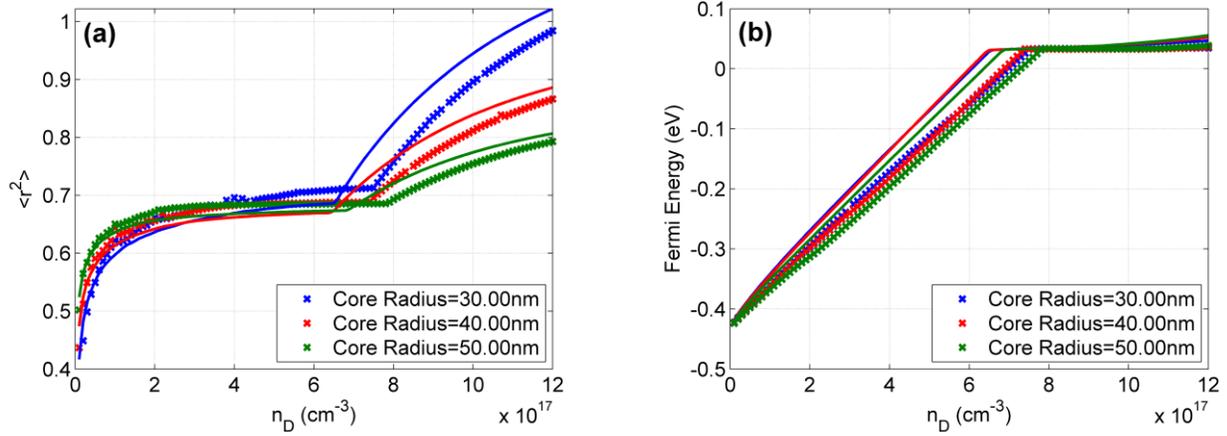

**Figure 6.** (a) Core-normalized variance of the electron density and (b) Fermi energy as a function of doping density, $n_D$. Calculations with exact exchange are denoted by crosses (x) and simulations without exact exchange are shown with solid lines.

We can understand these surprising trends in both $\langle r^2 \rangle$ and the Fermi Energy by using the PAMELA program to plot the conduction band and electron density for the nanowire near this critical density. Fig. 7a depicts the band-bending, Fermi energy, and electron density for a nanowire with a 30-nm-radius core and a 20-nm-thick shell with a doping density of $7.5 \times 10^{17}$ cm$^{-3}$. This specific doping density corresponds to the critical density for the 30 nm-radius core nanowire when exact exchange is included (blue crosses in Fig. 6a). In Fig. 7a, the band diagram shows that the Fermi energy is positioned well above the conduction band at the outer edge of the nanowire. For doping densities beyond this critical value as shown in Fig. 7b, tunneling into the shell layer occurs, and we observe some localization of the electron gas at the outer edge of the AlGaN shell. It is at this doping density that a sharp transition in $\langle r^2 \rangle$ begins to develop in Fig. 6a, and as the doping density is increased further, $\langle r^2 \rangle$ even exceeds 1 as the electron gas moves to the outer shell edge of the nanowire. It is interesting to note that we only observe these effects when the Fermi energy has reached a level where the shell barrier becomes thin enough for tunneling to occur. This further implies that there is a critical tunneling width (of the shell), that once reached, allows the electron gas to tunnel from the core into the shell layer. In all of our self-consistent calculations, we observe the onset of tunneling to occur when the width of the barrier is approximately two-thirds of the shell thickness, regardless of core radius.

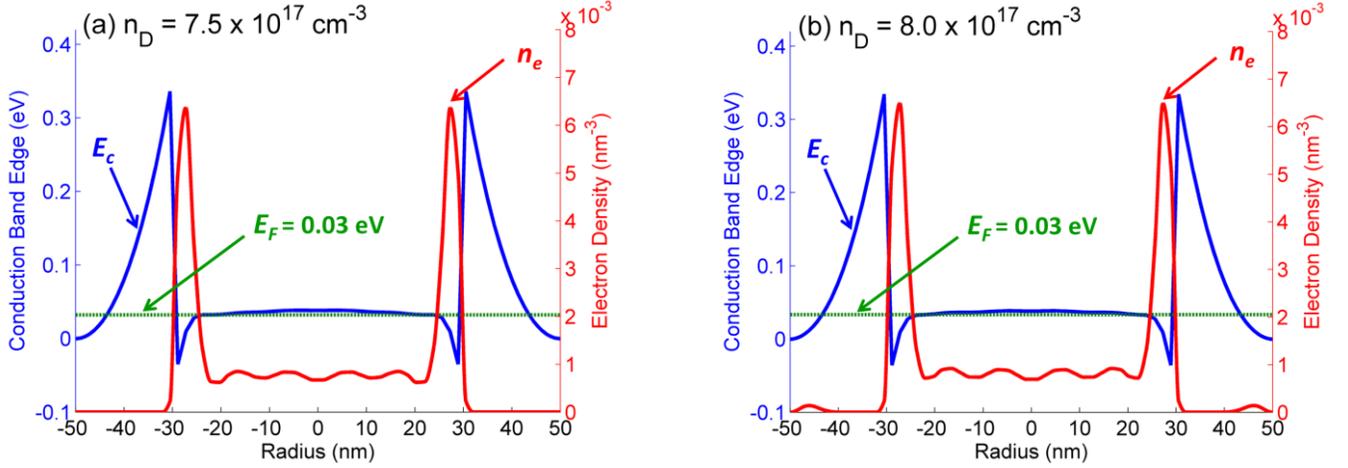

**Figure 7.** Band-bending (blue, left axis), Fermi energy (green), and electron density (red, right axis) for a nanowire with a 30-nm-radius core and a 20-nm-thick shell with doping density (a) $7.5 \times 10^{17}$ cm$^{-3}$ and (b) $8.0 \times 10^{17}$ cm$^{-3}$. At the higher doping density of $8.0 \times 10^{17}$ cm$^{-3}$, the free electron gas can be found at the outer edge of the Al$_{0.3}$Ga$_{0.7}$N shell. Both calculations for (a) and (b) were performed with nonlocal exact exchange.

**IV. Conclusions**

We have both demonstrated and provided an efficient approach for calculating the effects of nonlocal exact exchange on heterojunction electron gases in cylindrical core-shell nanowires. Our approach utilizes a pseudospectral numerical method that is efficiently implemented in the user-friendly PAMELA software package available online.[20] The extensive documentation and availability of our software package is a unique feature of this work since it provides a self-contained set of tools that both experimentalists and theorists can use to explore the effects of bandgap alignment, material composition, cross-sectional size, doping density, and many-body nonlocal exchange on electron gas formation in these core-shell nanowires. Furthermore, the availability of the fully open-source

PAMELA computer code encourages researchers to get a detailed "look under the hood" to obtain a deeper understanding of how these effects are numerically incorporated in practice. Using the PAMELA software package, we find that the inclusion of exact exchange has a crucial effect on the localization of electrons in a GaN/AlGaN core/shell nanowire and leads to a much larger energy separation between individual electronic subbands compared to simplistic treatments without exact exchange. In addition, we also investigated nanowire systems in the high doping density regime and found that tunneling from the core-shell interface to the nanowire edge becomes the dominant mechanism of electron gas formation. With the ability to efficiently carry out calculations with the PAMELA program, this work plays an important role in understanding which geometries and material parameters favor the spontaneous formation of electron gases in these nanoscale systems. Looking forward, it would be extremely interesting to understand the effects of exact exchange on both different cross-sectional geometries (i.e., hexagonal and triangular cross-sections[11]) and excited-state excitonic properties. Since electron/hole energies are extremely sensitive to the energy separation between electronic subbands, nonlocal exchange effects are expected to be essential for accurately predicting these excitations. These studies are currently underway in our group and will be presented in due course.

**Acknowledgement.** Funding for this effort was provided by the Laboratory Directed Research and Development (LDRD) program at Sandia National Laboratories, a multiprogram laboratory operated by Sandia Corporation, a Lockheed Martin Company, for the United States Department of Energy under contract DEAC04-94AL85000.

**Appendix**

Within this appendix and its subsections, we extensively document and give further details on the various approaches and numerical routines which comprise the PAMELA software package.[20] The source codes for PAMELA are written as simple MATLAB *m*-file programs that can calculate electronic energies, densities, wavefunctions, and band-bending diagrams within a self-consistent Schrödinger-Poisson formalism. Depending on the user input, the PAMELA program can calculate electronic properties incorporating either nonlocal exact exchange effects or just local direct interactions (i.e. a conventional Schrödinger-Poisson formalism). The self-contained PAMELA software package is freely available at http://alum.mit.edu/www/usagi.

**1. Schrödinger Equation**

Assuming translational invariance along the nanowire *z*-axis, the two-dimensional Schrödinger equation in the effective-mass approximation (c.f. Eq. (11)) is given by

$$\left[ -\frac{\hbar^2}{2} \nabla \cdot \frac{1}{m^*(r)} \nabla + V_n(r) + V_{CB}(r) + \sum_{j=1}^{N/2} 2V_{D,j}(r,\theta) - \sum_{j=1}^{N/2} V_{EXX,j}(r,\theta) \right] \psi_i(r,\theta) = E_i \psi_i(r,\theta), \quad (A1)$$

where $m^*$ is the effective mass, $\psi_i$ is the electron wavefunction for state *i*, $E_i$ is its energy, $V_n$ is the spatially dependent free carrier concentration due to the *n*-type doping density, $V_{CB}$ is the conduction band edge, $V_D$ is the (local) potential of the direct interaction, and $V_{EXX}$ is the (nonlocal) potential due to exact exchange. In the PAMELA source code for the Schrödinger equation, we make use of the dimensionless variables $\tilde{r}$ and $\varepsilon$ defined by the relations $r = \ell_0 \tilde{r}$ and $E = \varepsilon C$. Here, $C = \hbar^2 / (2 m_0 \ell_0^2)$, $m_0$ is the mass of an electron, and $\ell_0$ is a natural length scale that we define to be 1 nm. In terms of these dimensionless variables, Eq. (A1) reduces to:

$$\left\{ -\tilde{\nabla} \cdot \frac{1}{m^*(\tilde{r})/m_0} \tilde{\nabla} + \left[ V_{\text{CB}}(\tilde{r}) + \sum_{j=1}^{N/2} 2V_{\text{D},j}(\tilde{r},\theta) - \sum_{j=1}^{N/2} V_{\text{EXX},j}(\tilde{r},\theta) \right] \bigg/ C \right\} \psi_i(\tilde{r},\theta) = \varepsilon_i \psi_i(\tilde{r},\theta), \quad \text{(A2)}$$

where the kinetic energy term in (dimensionless) cylindrical polar coordinates is

$$\tilde{\nabla} \cdot \frac{1}{m^*(\tilde{r})/m_0} \tilde{\nabla} = \frac{\partial}{\partial \tilde{r}} \left[ \frac{1}{m^*(\tilde{r})/m_0} \right] \frac{\partial}{\partial \tilde{r}} + \frac{1}{\tilde{r} \cdot m^*(\tilde{r})/m_0} \frac{\partial}{\partial \tilde{r}} + \frac{1}{\tilde{r}^2 \cdot m^*(\tilde{r})/m_0} \frac{\partial^2}{\partial \theta^2}. \quad \text{(A3)}$$

For a cylindrical core-shell nanowire, both $m^*(\tilde{r})$ and $V_{\text{CB}}(\tilde{r})$ depend only on the radius $\tilde{r}$ and are coded in the PAMELA program as

$$\frac{1}{m^*(\tilde{r})/m_0} = \frac{1}{m^*_{\text{shell}}(\tilde{r})/m_0} + \left[ \frac{1}{m^*_{\text{core}}(\tilde{r})/m_0} - \frac{1}{m^*_{\text{shell}}(\tilde{r})/m_0} \right] \Theta(\tilde{r}_{\text{core}} - \tilde{r}), \quad \text{(A4)}$$

$$V_{\text{CB}}(\tilde{r}) = \Delta E_c \left[ \Theta(\tilde{r} - \tilde{r}_{\text{core}}) - 1 \right], \quad \text{(A5)}$$

where $m^*_{\text{core}}$ is the effective mass of the core, $m^*_{\text{shell}}$ is the effective mass of the shell, $\tilde{r}_{\text{core}}$ is the radius of the core region, $\Delta E_c$ is the conduction band offset, and $\Theta(\tilde{r})$ (= 0 for $\tilde{r} < 0$ or = 1 for $\tilde{r} > 0$) is the conventional Heaviside step function. Note that for $\tilde{r} < \tilde{r}_{\text{core}}$, Eq. (A4) and Eq. (A5) reduce to $1/\left[ m^*(\tilde{r})/m_0 \right] = 1/\left[ m^*_{\text{core}}(\tilde{r})/m_0 \right]$ and $V_{\text{CB}} = -\Delta E_c$, respectively; conversely for $\tilde{r} > \tilde{r}_{\text{core}}$, Eq. (A4) and Eq. (A5) reduce to $1/\left[ m^*(\tilde{r})/m_0 \right] = 1/\left[ m^*_{\text{shell}}(\tilde{r})/m_0 \right]$ and $V_{\text{CB}} = 0$.

As required in quantum mechanics, the *unscaled* wavefunctions, $\psi_i(r,\theta)$, must be normalized over the entire nanowire cross-section; i.e., in polar coordinates, $\int dr \, d\theta \, r |\psi_i(r,\theta)|^2 = 1$. However, in the PAMELA source code, the wavefunctions in dimensionless coordinates, $\psi_i(\tilde{r},\theta)$, are normalized *in the reduced coordinate $\tilde{r}$* and $\theta$; i.e., $\int d\tilde{r} \, d\theta \, \tilde{r} |\psi_i(\tilde{r},\theta)|^2 = 1$. In order to satisfy both of these constraints, we must have the definition

$$|\psi(r,\theta)|^2 \equiv |\psi(\tilde{r},\theta)|^2 / \ell_0^2 \quad \text{(A6)}$$

As discussed briefly in Chapter 9 of Ref. 18, the pseudospectral method computes eigenfunctions that are linearly independent but not necessarily orthogonal to each other. Since we require the wavefunctions to be mutually orthogonal, we use Löwdin's procedure (also known as symmetric orthogonalization[33]) in the PAMELA program to orthogonalize our wavefunctions. This procedure is carried out by constructing the following $N/2 \times N/2$ overlap matrix:

$$\mathbf{S} = \begin{pmatrix} S_{1,1} & S_{1,2} & \cdots & S_{1,N/2} \\ S_{2,1} & S_{2,2} & & \\ \vdots & & \ddots & \vdots \\ S_{N/2,1} & & \cdots & S_{N/2,N/2} \end{pmatrix}, \quad (A7)$$

where each matrix element, $S_{i,j}$, is given by the overlap integral

$$S_{i,j} = \int d\tilde{r}\, d\theta\, \tilde{r}\, \psi_i^*(\tilde{r},\theta)\psi_j(\tilde{r},\theta), \quad (A8)$$

and $N/2$ is the number of doubly-occupied spatial orbitals with energies below the Fermi energy (see Section 4 of the Appendix). The inverse square root of $\mathbf{S}$ forms a transformation matrix $\mathbf{S}^{-1/2}$ such that an orthogonal set of wavefunctions, $\{\psi_j'\}$, is given by

$$\psi_j' = \sum_i \psi_i \cdot \left(\mathbf{S}^{-1/2}\right)_{i,j}, \quad (A9)$$

where $\{\psi_i\}$ is the set of non-orthogonal wavefunctions obtained from the pseudospectral method. Within the PAMELA source code, the set of wavefunctions $\{\psi_i\}$ and $\{\psi_j'\}$ are actually contained in column-matrices, and Eq. (A9) can be expressed in matrix notation as

$$\mathbf{\psi}' = \mathbf{\psi} \cdot \mathbf{S}^{-1/2}, \quad (A10)$$

where $\mathbf{\psi}'$ and $\mathbf{\psi}$ are column-matrices of orthogonal and non-orthogonal wavefunctions, respectively.

**2. $V_n$ and the Direct Interaction**

Within the PAMELA source code, both $V_n$ and $V_D$ are solved simultaneously in a single equation (both $V_n$ and $V_D$ are local potentials):

$$\nabla \cdot \varepsilon(r)\varepsilon_0 \nabla \left[ V_n(r,\theta) + \sum_{j=1}^{N/2} 2V_{D,j}(r,\theta) \right] = |e|\rho_D(r,\theta) - 2e^2 \sum_{j=1}^{N/2} |\psi'_j(r,\theta)|^2 N(E_F, E_j, T), \quad (A11)$$

where $\psi'_j$ are orthogonal wavefunctions (see end of Section 1 in the Appendix), and we have made use of Eqs. (16) and (13). The integrated electron concentration is given by

$$2N(E_F, E_j, T) = \int_{E_j}^{\infty} dE\, g(E, E_j) f(E, E_F, T)$$
$$= \int_{E_j}^{\infty} dE\, \sqrt{\frac{m^*(r)}{2\pi^2\hbar^2(E-E_j)}} f(E, E_F, T), \quad (A12)$$

where $g(E, E_j)$ is the one-dimensional density of states (Eq. (A12) also takes account of the spin degeneracy of electrons), $E_F$ is the Fermi energy (described further in Section 4 of the Appendix), $T$ is the temperature, and

$$f(E, E_F, T) = \begin{cases} \dfrac{1}{e^{(E-E_F)/k_B T} + 1}, & \text{for } T \neq 0, \\ \Theta(E_F - E), & \text{for } T = 0, \end{cases} \quad (A13)$$

is the Fermi-Dirac distribution. For $T = 0$ K, which is the only case implemented in the PAMELA program, the integral in Eq. (A12) can be evaluated analytically and reduces to

$$2N(E_F, E_j, T) = \int_{E_j}^{\infty} dE\, \sqrt{\frac{m^*(r)}{2\pi^2\hbar^2(E-E_j)}} \Theta(E_F - E)$$
$$= \int_{E_j}^{E_F} dE\, \sqrt{\frac{m^*(r)}{2\pi^2\hbar^2(E-E_j)}} \quad (A14)$$
$$= \sqrt{\frac{2m^*(r)(E_F - E_j)}{\pi^2 \hbar^2}}.$$

It is also useful to express the charge density, $\rho_D$, in Eq. (A11) in terms of a number density, $n_D$:

$$\rho_D(r,\theta) = |e| n_D(r,\theta). \quad (A15)$$

Using Eqs. (A14) and (A15), Eq. (A11) reduces to

$$\nabla \cdot \varepsilon(r) \nabla \left[ V_n(r,\theta) + \sum_{j=1}^{N/2} 2V_{D,j}(r,\theta) \right] = \frac{e^2}{\varepsilon_0} \left[ n_D(r,\theta) - \sum_{j=1}^{N/2} |\psi'_j(r,\theta)|^2 \sqrt{\frac{2m^*(r)(E_F - E_j)}{\pi^2 \hbar^2}} \right]. \quad (A16)$$

In terms of the dimensionless variables $\tilde{r}$ and $\varepsilon$ defined by the relations $r = \ell_0 \tilde{r}$ and $E = \varepsilon C$, Eq. (A16) reduces to:

$$\tilde{\nabla} \cdot \varepsilon(\tilde{r}) \tilde{\nabla} \left[ V_n(\tilde{r},\theta) + \sum_{j=1}^{N/2} 2V_{D,j}(\tilde{r},\theta) \right]$$
$$= \frac{\ell_0^2 e^2}{\varepsilon_0} \left[ n_D(\tilde{r},\theta) - \frac{1}{\pi \ell_0^3} \sum_{j=1}^{N/2} |\psi'_j(\tilde{r},\theta)|^2 \sqrt{[m^*(\tilde{r})/m_0](\varepsilon_F - \varepsilon_j)} \right] \quad (A17)$$
$$= (18.0951 \text{ eV} \cdot \text{nm}^3) \left[ n_D(\tilde{r},\theta) - \frac{1}{\pi \ell_0^3} \sum_{j=1}^{N/2} |\psi'_j(\tilde{r},\theta)|^2 \sqrt{[m^*(\tilde{r})/m_0](\varepsilon_F - \varepsilon_j)} \right],$$

where we have made use of the definition in Eq. (A6). Within the PAMELA program, our number density $n_D$ is assumed to be constant and is inputted in units of $1/\ell_0^3$ ($= 1 \text{ nm}^{-3}$). The differential operator in (dimensionless) cylindrical polar coordinates is

$$\tilde{\nabla} \cdot \varepsilon(\tilde{r}) \tilde{\nabla} = \frac{\partial}{\partial \tilde{r}} \left[ \varepsilon(\tilde{r}) \right] \frac{\partial}{\partial \tilde{r}} + \frac{\varepsilon(\tilde{r})}{\tilde{r}} \frac{\partial}{\partial \tilde{r}} + \frac{\varepsilon(\tilde{r})}{\tilde{r}^2} \frac{\partial^2}{\partial \theta^2}. \quad (A18)$$

The position-dependent dielectric $\varepsilon(\tilde{r})$ in a cylindrical core-shell nanowire depends only on the radius $\tilde{r}$ and is coded in the PAMELA program as

$$\varepsilon(\tilde{r}) = \varepsilon_{\text{shell}}(\tilde{r}) + \left[ \varepsilon_{\text{core}}(\tilde{r}) - \varepsilon_{\text{shell}}(\tilde{r}) \right] \Theta(\tilde{r}_{\text{core}} - \tilde{r}), \quad (A19)$$

where $\varepsilon_{\text{core}}$ is the dielectric of the core and $\varepsilon_{\text{shell}}$ is the dielectric of the shell. Again, note that for $\tilde{r} < \tilde{r}_{\text{core}}$, Eq. (A19) reduces to $\varepsilon(\tilde{r}) = \varepsilon_{\text{core}}(\tilde{r})$, and for $\tilde{r} > \tilde{r}_{\text{core}}$, Eq. (A19) becomes $\varepsilon(\tilde{r}) = \varepsilon_{\text{shell}}(\tilde{r})$.

### 3. The Exchange Interaction

From Eq. (15), the exchange interaction, $V_{\text{EXX}}$, is implemented within the PAMELA source code as

$$\nabla \cdot \varepsilon(r)\varepsilon_0 \nabla \left[ \sum_{j=1}^{N/2} V_{\text{EXX},j}(r,\theta) \right] = -e^2 \sum_{j=1}^{N/2} \psi_j'^*(r,\theta)\psi_i'(r,\theta) N(E_F, E_j, T). \tag{A20}$$

Using Eq. (A14) for the integrated electron concentration, $N(E_F, E_j, T)$, Eq. (A20) reduces to

$$\nabla \cdot \varepsilon(r) \nabla \left[ \sum_{j=1}^{N/2} V_{\text{EXX},j}(r,\theta) \right] = -\frac{e^2}{2\varepsilon_0} \sum_{j=1}^{N/2} \psi_j'^*(r,\theta)\psi_i'(r,\theta) \sqrt{\frac{2m^*(r)(E_F - E_j)}{\pi^2 \hbar^2}}. \tag{A21}$$

In terms of the dimensionless variables $\tilde{r}$ and $\varepsilon$, Eq. (A21) reduces to:

$$\tilde{\nabla} \cdot \varepsilon(\tilde{r}) \tilde{\nabla} \left[ \sum_{j=1}^{N/2} V_{\text{EXX},j}(\tilde{r},\theta) \right]$$
$$= -\frac{\ell_0^2 e^2}{\varepsilon_0} \left[ \frac{1}{2\pi \ell_0^3} \sum_{j=1}^{N/2} \psi_j'^*(r,\theta)\psi_i'(r,\theta) \sqrt{\left[m^*(\tilde{r})/m_0\right](\varepsilon_F - \varepsilon_j)} \right] \tag{A22}$$
$$= -(18.0951 \text{ eV} \cdot \text{nm}^3) \left[ \frac{1}{2\pi \ell_0^3} \sum_{j=1}^{N/2} \psi_j'^*(\tilde{r},\theta)\psi_i'(\tilde{r},\theta) \sqrt{\left[m^*(\tilde{r})/m_0\right](\varepsilon_F - \varepsilon_j)} \right],$$

where we have again made use of the definition in Eq. (A6). The differential operator and position-dependent dielectric $\varepsilon(\tilde{r})$ in dimensionless cylindrical polar coordinates are given by Eqs. (A18) and (A19), respectively.

### 4. The Fermi Energy and Charge Neutrality

We determine the Fermi level, $E_F$, by requiring the total number of positive and negative charges over the entire nanowire to be equal:

$$\int d\tilde{r} \, d\theta \, \tilde{r} \, n_D(\tilde{r},\theta) = \int d\tilde{r} \, d\theta \, \tilde{r} \, n_e(\tilde{r},\theta), \tag{A23}$$

where $n_e(\tilde{r},\theta)$ in dimensionless variables is given by

$$n_e(\tilde{r},\theta) = \frac{1}{\pi \ell_0^3} \sum_{j=1}^{N/2} |\psi_j'(\tilde{r},\theta)|^2 \sqrt{\left[m^*(\tilde{r})/m_0\right](\varepsilon_F - \varepsilon_j)} \tag{A24}$$

In the PAMELA source code, we assume a uniform doping density of $n_D$ in both the core and shell regions, and Eq. (A23) reduces to

$$\int d\tilde{r}\, d\theta\, \tilde{r}\, n_e(\tilde{r},\theta) - n_D \pi \tilde{r}_{shell}^2 = 0, \tag{A25}$$

where $\tilde{r}_{shell}$ is the radius of the outer shell region. Since $n_e$ is a function of $E_F$ (see Eq. (A24)), the Fermi energy is determined by solving Eq. (A25) with a standard root finding algorithm within PAMELA.

### 5. Boundary Conditions and Iteration Procedure

As discussed in the Theory and Methodology Section, we assume Dirichlet boundary conditions for the Schrödinger, direct interaction, and exchange interaction equations where the wavefunctions and various interaction potentials are set to zero at the outer shell boundary. The PAMELA program converts the dimensionless partial differential expressions in Eqs. (A2), (A17), and (A22) to matrix equations using pseudospectral methods (discussed further in Sections 6-9 of the Appendix). For the Schrödinger equation in Eq. (A2), the entire expression within the curly braces can be cast into a single (dense) matrix. The solution to Eq. (A2) using pseudospectral methods then amounts to finding the eigenvalues of this matrix, with the eigenvectors forming a numerical representation of the wavefunctions at specific grid points.[18] For Eqs. (A17), and (A22), the differential operator $\tilde{\nabla} \cdot \varepsilon(\tilde{r}) \tilde{\nabla}$ is also a dense matrix that can be inverted numerically after boundary conditions are specified. The pseudospectral solution for the direct interaction in Eq. (A17) has the form

$$\left[ V_n(\tilde{r},\theta) + \sum_{j=1}^{N/2} 2V_{D,j}(\tilde{r},\theta) \right]$$
$$= (18.0951\ \text{eV}\cdot\text{nm}^3) \left\{ \left[ \tilde{\nabla} \cdot \varepsilon(\tilde{r}) \tilde{\nabla} \right]^{-1} \left[ n_D(\tilde{r},\theta) - \frac{1}{\pi \ell_0^3} \sum_{j=1}^{N/2} |\psi_j'(\tilde{r},\theta)|^2 \sqrt{[m^*(\tilde{r})/m_0](\varepsilon_F - \varepsilon_j)} \right] \right\}, \tag{A26}$$

where $\left[\tilde{\nabla}\cdot\varepsilon(\tilde{r})\tilde{\nabla}\right]^{-1}$ denotes the matrix inverse of the differential operator matrix. The effect of the direct interaction (Eq. (A26)) on a wavefunction $\psi'_i(\tilde{r},\theta)$ in the Schrödinger equation is defined by Eq. (8) and is implemented in the PAMELA source code as

$$\left[V_n(\tilde{r},\theta)+\sum_{j=1}^{N/2}2V_{D,j}(\tilde{r},\theta)\right]\psi'_i(\tilde{r},\theta)$$
$$=(18.0951\text{ eV}\cdot\text{nm}^3)\left\{\left[\tilde{\nabla}\cdot\varepsilon(\tilde{r})\tilde{\nabla}\right]^{-1}\left[n_D(\tilde{r},\theta)-\frac{1}{\pi\ell_0^3}\sum_{j=1}^{N/2}\left|\psi'_j(\tilde{r},\theta)\right|^2\sqrt{\left[m^*(\tilde{r})/m_0\right](\varepsilon_F-\varepsilon_j)}\right]\right\}\psi'_i(\tilde{r},\theta),$$
(A27)

In Eq. (A27), the inverse matrix operator $\left[\tilde{\nabla}\cdot\varepsilon(\tilde{r})\tilde{\nabla}\right]^{-1}$ operates only within the curly braces; that is, the direct interaction is a local (multiplicative) potential function that essentially multiplies the wavefunction $\psi'_i(\tilde{r},\theta)$ in the Schrödinger equation. The pseudospectral expression for the exchange interaction in Eq. (A22) has the form

$$\left[\sum_{j=1}^{N/2}V_{\text{EXX},j}(\tilde{r},\theta)\right]$$
$$=-(18.0951\text{ eV}\cdot\text{nm}^3)\left\{\left[\tilde{\nabla}\cdot\varepsilon(\tilde{r})\tilde{\nabla}\right]^{-1}\left[\frac{1}{2\pi\ell_0^3}\sum_{j=1}^{N/2}\psi'^*_j(\tilde{r},\theta)\psi'_i(\tilde{r},\theta)\sqrt{\left[m^*(\tilde{r})/m_0\right](\varepsilon_F-\varepsilon_j)}\right]\right\},$$
(A28)

The effect of the exchange interaction (Eq. (A28)) on a wavefunction $\psi'_i(\tilde{r},\theta)$ is defined by Eq. (9) and is more complicated:

$$\left[\sum_{j=1}^{N/2}V_{\text{EXX},j}(\tilde{r},\theta)\right]\psi'_i(\tilde{r},\theta)$$
$$=-(18.0951\text{ eV}\cdot\text{nm}^3)\psi'_j(\tilde{r},\theta)\left\{\left[\tilde{\nabla}\cdot\varepsilon(\tilde{r})\tilde{\nabla}\right]^{-1}\left[\frac{1}{2\pi\ell_0^3}\sum_{j=1}^{N/2}\psi'^*_j(\tilde{r},\theta)\psi'_i(\tilde{r},\theta)\sqrt{\left[m^*(\tilde{r})/m_0\right](\varepsilon_F-\varepsilon_j)}\right]\right\},$$
(A29)

where the inverse matrix operator $\left[\tilde{\nabla}\cdot\varepsilon(\tilde{r})\tilde{\nabla}\right]^{-1}$ again operates only within the curly braces. In contrast to the direct interaction, note that the inverse matrix operator *does* operate on the wavefunction $\psi'_i(\tilde{r},\theta)$ in the Schrödinger equation, and the exchange interaction is a nonlocal, integral operator (the inverse matrix operator is mathematically equivalent to an integral operator).

The coupled set of Schrödinger, direct, and exchange interaction equations in Eqs. (A2), (A27), and (A29) are solved iteratively until self-consistency is achieved. For our starting guess to initialize the self-consistent procedure in the PAMELA source code, the bare conduction band offset (Eq. (A5)) for the core-shell system is used as a seed potential for the Schrödinger equation to generate a set of normalized wavefunctions and eigenenergies. The wavefunctions are orthogonalized using the Löwdin orthogonalization procedure, and these solutions are then subsequently used in the calculation of the Poisson and Exchange interaction potentials. The resulting potentials are then incorporated into a new Schrödinger equation for the next iteration, and the entire process iterates until convergence. To maintain a stable cycle for iteration to self-consistency, we use an under-relaxation technique on the matrix representation of the total interaction potential with an under-relaxation parameter $\omega$ (set to 0.05 in the PAMELA source code) such that after iteration $i$, the matrix representation of the total input potential ($\mathbf{V}_{\text{input},i}$) is calculated as

$$\mathbf{V}_{\text{input},i} = \mathbf{V}_{\text{input},i-1} + \omega\left(\mathbf{V}_n + \mathbf{V}_{\text{CB}} + 2\mathbf{V}_{\text{D},i} - \mathbf{V}_{\text{EXX},i} - \mathbf{V}_{\text{input},i-1}\right). \tag{A30}$$

These iterations are continued to self-consistency, which we have defined as the maximum change among all the matrix elements of the full interaction potential between successive iterations to fall below 0.001 eV.

## 6. Pseudospectral Methods

As discussed extensively in Ref. 18, pseudospectral methods are an efficient and accurate approach for numerically solving partial differential equations and computing integrals. In contrast to finite difference or finite element methods that approximate a differential operator by a large and sparse matrix (i.e., first- or second-order differentiation schemes), the idea behind pseudospectral methods is to utilize a differentiation matrix of infinite order and infinite bandwidth, leading to a dense matrix. Although the expressions for pseudospectral differentiation matrices can be more complicated, one can achieve significantly improved accuracy and speed compared to finite difference or finite element schemes (for instance, the first 2 computer programs in Ref. 18 show a striking example where a fourth-order finite difference method requires a $\sim 10^3 \times 10^3$ matrix to achieve the same convergence as a much smaller $\sim 10 \times 10$ matrix using the pseudospectral method). These results can be generalized to multiple dimensions, which we briefly review in Section 9 of the Appendix. While we present the pseudospectral equations required to understand the inner workings of the PAMELA source code, the reader is strongly encouraged to thoroughly review Ref. 18 for a more detailed explanation and implementation of pseudospectral methods.

## 7. Periodic functions

The most natural grid to use for differential equations with one-dimensional periodic boundary conditions is an equispaced grid over the interval $[0, 2\pi]$. For $N$ grid points within this interval, the spacing between adjacent grid points $\theta_j$ ($j = 1,\ldots,N$) is $h = 2\pi/N$. Following the procedure and notation of Ref. 18, we can obtain an expression for the second derivative of the band-limited interpolant of the periodic delta function:

$$S_N''(\theta_j) = \begin{cases} -\frac{\pi^2}{3h^2} - \frac{1}{6}, & j = 0 \pmod{N}, \\ -\frac{(-1)^j}{2\sin^2(jh/2)}, & j \neq 0 \pmod{N}. \end{cases} \quad (A31)$$

As discussed in Ref. 18, these numbers form the entries of the $N$th column of the $N \times N$ second-order spectral differentiation matrix. Specifically, the second-order spectral differentiation matrix can be written as a Toeplitz matrix:

$$D_\theta^{(2)} = \begin{pmatrix} \ddots & & & \vdots & & & \\ \ddots & -\frac{1}{2}\csc^2\left(\frac{2h}{2}\right) & & & & & \\ \ddots & & \frac{1}{2}\csc^2\left(\frac{h}{2}\right) & & & & \\ & & & -\frac{\pi^2}{3h^2} - \frac{1}{6} & & & \\ & & & & \frac{1}{2}\csc^2\left(\frac{h}{2}\right) & \ddots & \\ & & & & -\frac{1}{2}\csc^2\left(\frac{2h}{2}\right) & & \ddots \\ & & & \vdots & & & \ddots \end{pmatrix} \tag{A32}$$

It is important to mention at this point that Eqs. (A31) and (A32) are only valid *if the number of grid points, N, in the periodic grid is even*. There are analogous expressions for odd *N*, but the formulas are different. In order to keep the PAMELA source code as simple as possible, *the number of angular grid points, N, inputted into the* PAMELA *program must be even*.

In order to calculate integrals over periodic domains with spectral accuracy, the periodic Fourier integration of a function $f(\theta)$ reduces to the periodic trapezoidal rule:

$$\int_0^{2\pi} d\theta\, f(\theta) = \frac{2\pi}{N} \sum_{j=1}^{N} f(\theta_j), \tag{A33}$$

or in matrix form,

$$\int_0^{2\pi} d\theta\, f(\theta) = \frac{2\pi}{N} \mathbf{w}^\mathrm{T} \cdot \mathbf{f}, \tag{A34}$$

where $\mathbf{w}^\mathrm{T}$ is a weight vector with $N$ elements: $(1, 1, 1, \ldots, 1)^\mathrm{T}$, and $\mathbf{f}$ is a row vector with entries $(f(\theta_1), f(\theta_2), f(\theta_3), \ldots, f(\theta_N))$.

## 8. Non-Periodic functions

For differential equations defined over a one-dimensional interval that is non-periodic, we can obtain spectral accuracy by using a Chebychev discretization given by the following (non-equispaced) grid points:

$$r_j = \cos\left(\frac{j\pi}{N}\right), \quad j = 0, 1, \ldots, N. \tag{A35}$$

Following the procedure and notation of Ref. 18, the entries of the $N \times N$, first-order Chebyshev spectral differential matrix are

$$\left(D_r^{(1)}\right)_{0,0} = \frac{2N^2 + 1}{6}, \quad \left(D_r^{(1)}\right)_{N,N} = -\frac{2N^2 + 1}{6}, \tag{A36}$$

$$\left(D_r^{(1)}\right)_{j,j} = -\frac{r_j}{2(1 - r_j^2)}, \quad j = 1, \ldots, N - 1, \tag{A37}$$

$$\left(D_r^{(1)}\right)_{i,j} = \frac{c_i}{c_j} \frac{(-1)^{i+j}}{(r_i - r_j)}, \quad i \neq j, \quad i, j = 1, \ldots, N - 1, \tag{A38}$$

where

$$c_i = \begin{cases} 2 & i = 0 \text{ or } N, \\ 1 & \text{otherwise.} \end{cases} \tag{A39}$$

Higher-order differentiation matrices can be generated by taking successive powers of $D_r^{(1)}$. The generation of this matrix is implemented in Trefethen's MATLAB function, cheb.m,[18] which is included in the PAMELA software package. Note, that the choice of Chebyshev grid points in Eq. (A35) restricts us to differential equations that are only defined within the interval [-1, 1]. Since our cylindrical nanowire is defined within the domain $|\tilde{r}| \leq \tilde{r}_{\text{shell}}$, we must rescale our coordinates and the differentiation matrix $\left(D_r^{(1)}\right)$ to the [-1, 1] Chebyshev domain. Within the beginning of the PAMELA source code, we implement rescaled coordinates by defining the variables:

$$\tilde{r} = \tilde{r}_{\text{shell}} \tilde{r}_s, \tag{A40}$$

and

$$\frac{\partial}{\partial \tilde{r}} = \frac{1}{\tilde{r}_{\text{shell}}} \frac{\partial}{\partial \tilde{r}_s}, \tag{A41}$$

where $\tilde{r}$ ranges from $-\tilde{r}_{shell}$ to $\tilde{r}_{shell}$, while $\tilde{r}_s$ ranges from -1 to 1. The definitions in Eqs. (A40) and (A41) are substituted into the expressions for the differential operators in Eqs. (A3) and (A18), which are later converted into differentiation matrices in the PAMELA source code.

In order to calculate integrals over non-periodic domains with spectral accuracy, we utilize Clenshaw-Curtis quadrature to evaluate integrals of the form

$$\int_{-1}^{1} dr_s\, f(r_s) = \mathbf{w}^\mathbf{T} \cdot \mathbf{f}, \tag{A42}$$

where $\mathbf{w}^\mathbf{T}$ is a weight vector computed with Trefethen's MATLAB function, clencurt.m,[18] which is also included in the PAMELA software package, and $\mathbf{f}$ is a row vector with entries $(f(r_1), f(r_2), f(r_3), \ldots, f(r_N))$.

## 9. 2-Dimensional Grids and Differentiation Matrices

Since the coupled differential equations for our cylindrical nanowire depend on both $\tilde{r}$ and $\theta$, we must construct all of the various operators in two dimensions. For example, if we have the pseudospectral differential matrices $\mathbf{D}_\mathbf{r}^{(2)}$ and $\mathbf{D}_\theta^{(2)}$ of dimensions $N \times N$ and $M \times M$, respectively, we can construct

$$\mathbf{D}_{\mathbf{r},2D}^{(2)} = \mathbf{D}_\mathbf{r}^{(2)} \otimes \mathbf{I}_M, \quad \mathbf{D}_{\theta,2D}^{(2)} = \mathbf{I}_N \otimes \mathbf{D}_\theta^{(2)}, \tag{A43}$$

where $\mathbf{I}_N$ is the $N \times N$ identity matrix, $\mathbf{I}_M$ is the $M \times M$ identity matrix, and $\otimes$ denotes the Kronecker product. Both $\mathbf{D}_{\mathbf{r},2D}^{(2)}$ and $\mathbf{D}_{\theta,2D}^{(2)}$ have dimensions of $(N \times M) \times (N \times M)$.

We discretize the cross-section of the cylindrical nanowire by taking a periodic Fourier grid in $\theta \in [0, 2\pi]$ and a non-periodic Chebyshev grid in $r \in [-1, 1]$. This choice of discretization results in a mapping where each point $(x, y)$ in the circular cross-section corresponds to two distinct points $(r, \theta)$ in coordinate space; i.e., the map from $(r, \theta)$ to $(x, y)$ is 2-to-1. We can avoid this complication by first using an odd number of grid points, $N$, in the $r$ direction to avoid the $\infty$-to-1 mapping at the origin. Next, we can reduce the size of our matrices by taking advantage of the two-fold symmetry inherent in the matrix structure as discussed further in Chapter 11 of Ref. 18. Using these techniques, a two-fold reduction in complexity is attained, resulting in the set of discretized equations found within the PAMELA source code.[20]

**References**

[1]P. Yang, R. Yan, and M. Fardy, Nano Lett. **10**, 1529 (2010).

[2]L. J. Lauhon, M. S. Gudiksen, D. Wang, and C. M. Lieber, Nature **420**, 57 (2002).

[3]F. Qian, Y. Li, S. Gradečak, D. Wang, C. J. Barrelet, and C. M. Lieber, Nano Lett. **4**, 1975 (2004).

[4]W. Lu, J. Xiang, B. P. Timko, Y. Wu, and C. M. Lieber, Proc. Natl. Acad. Sci. U.S.A. **102**, 10046 (2005).

[5]Y. Li, J. Xiang, F. Qian, S. Gradečak, Y. Wu, H. Yan, D. A. Blom, and C. M. Lieber, Nano Lett. **6**, 1468 (2006).

[6]F. Qian, S. Gradečak, Y. Li, C.-Y. Wen, and C. M. Lieber, Nano Lett. **5**, 2287 (2005).

[7]N. Sköld, L. S. Karlsson, M. W. Larsson, M.-E. Pistol, W. Seifert, J. Trägårdh, and L. Samuelson, Nano Lett. **5**, 1943 (2005).

[8]O. Hayden, A. B. Greytak, and D. C. Bell, Adv. Mater. **17**, 701 (2005).

[9]K. Hestroffer, R. Mata, D. Camacho, C. Ledere, G. Tourbot, Y. M. Niquet, A. Cros, C. Bourgerol, H. Renevier, B. Daudin, Nanotechnology **21**, 415702 (2010).

[10]L. Wang, D. Wang, and P. M. Asbeck, Solid-State Electron. **50**, 1732 (2006).

[11]B. M. Wong, F. Léonard, Q. Li, and G. T. Wang, Nano Lett. **11**, 3074 (2011).

[12]A. Bertoni, M. Royo, F. Mahawish, G. Goldoni, Phys. Rev. B **84**, 205323 (2011).

[13]J. P. Perdew and A. Zunger, Phys. Rev. B **23**, 5048 (1981).

[14]J. P. Perdew and M. Levy, Phys. Rev. B **56**, 16021 (1997).

[15]Y. Zhang and W. Yang, J. Chem. Phys. **109**, 2604 (1998).

[16]U. Lundin and O. Eriksson, Int. J. Quant. Chem. **81**, 247 (2001).

[17]V. Polo, E. Kraka, and D. Cremer, Mol. Phys. **100**, 1771 (2002).

[18]L. N. Trefethen, Spectral Methods in MATLAB (SIAM, Philadelphia, 2000). See http://people.maths.ox.ac.uk/trefethen/spectral.html

[19]J. S. Heyl and A. Thirumalai, Mon. Not. R. Astron. Soc. **407**, 590 (2010).

[20]See http://alum.mit.edu/www/usagi

[21]B. M. Wong and J. G. Cordaro, J. Chem. Phys. **129**, 214703 (2008).

[22]B. M. Wong, M. Piacenza, F. D. Sala, Phys. Chem. Chem. Phys. **11**, 4498 (2009).

[23]B. M. Wong and T. H. Hsieh, J. Chem. Theory Comput. **6**, 3704 (2010).

[24]B. M. Wong and S. H. Ye, Phys. Rev. B **84**, 075115 (2011).

[25]B. M. Wong and J. G. Cordaro, J. Phys. Chem. C **115**, 18333 (2011).

[26]M. E. Foster and B. M. Wong, J. Chem. Theory Comput. **8**, 2682 (2012).

[27]T. Stein, L. Kronik, R. Baer, J. Am. Chem. Soc. **131**, 2818 (2009).

[28]L. Kronik, T. Stein, S. Refaely-Abramson, R. Baer, J. Chem. Theory Comput. **8**, 1515 (2012).

[29]C. Møller and M. S. Plesset, Phys. Rev. **46**, 618 (1934).

[30] L. Hedin, Phys. Rev. **139**, A796 (1965).

[31] M. S. Hybertsen and S. G. Louie, Phys. Rev. B **34**, 5390 (1986).

[32] K. Andersson, P. A. Malmqvist, B. O. Roos, A. J. Sadleg, and K. Wolinski, J. Phys. Chem. **94**, 5483 (1990).

[33] A. Szabo and N. S. Ostlund, Modern Quantum Chemistry: Introduction to Advanced Electronic Structure Theory (Dover, New York, 1989).

[34] A. Thirumalai and J. S. Heyl, Phys. Rev. A **79**, 012514 (2009).

[35] L. Laaksonen, P. Pyykkö, and D. Sundholm, Chem. Phys. Lett. **96**, 1 (1983).

[36] V. A. Fonoberov and A. A. Balandin, J. Appl. Phys. **94**, 7178 (2003).